# Deep learning for dynamic modeling and coded information storage of vector-soliton pulsations in mode-locked fiber lasers

Zhi-Zeng Si[1#], Da-Lei Wang[1#], Bo-Wei Zhu[1], Zhen-Tao Ju[1], Xue-Peng Wang[1], Wei Liu[1*], Boris A. Malomed[2,3], Yue-Yue Wang [1*] and Chao-Qing Dai[1*]

[1]*College of Optical, Mechanical and Electrical Engineering, Zhejiang A&F University, Lin'an 311300, China*

[2]*Department of Physical Electronics, Faculty of Engineering, and Center for Light-Matter Interaction, Tel Aviv University, Tel Aviv 69978, Israel*
[3]*Instituto de Alta Investigación, Universidad de Tarapacá, Casilla 7D, Arica 1000000, Chile*

**Abstract:** Soliton pulsations are ubiquitous feature of non-stationary soliton dynamics in mode-locked lasers and many other physical systems. To overcome difficulties related to huge amount of necessary computations and low efficiency of traditional numerical methods in modeling the evolution of non-stationary solitons, we propose a two-parallel bidirectional long short-term memory recurrent neural network (TP-Bi_LSTM RNN), with the main objective to predict dynamics of vector-soliton pulsations (VSPs) in various complex states, whose real-time dynamics is verified by experiments. For two examples, *viz*., single- and bi-periodic VSPs**,** with period-21 and a combination of period-3 and period-43, the prediction results are better than provided by direct simulations – namely, deviations produced by the RNN results are 36% and 18% less than those provided by the simulations, respectively. This means that predicted results provided by the neural network are better than numerical simulations. Moreover, the prediction results for unstable VSP state with period-9 indicate that the optimization of training sets and the number of training iterations are particularly important for the predictability. Besides, the scheme of coded information storage based on the TP-Bi_LSTM RNN, instead of actual pulse signals, is realized too. The findings offer new applications of deep learning to ultrafast optics and information storage.

## 1. Introduction

Soliton pulsations (SP)[1, 2] are a broad class of self-trapped localized modes with internal periodic oscillations determined by parameters of the physical system (in particular, of optical resonators)[3, 4]. SPs have been widely studied in fluid dynamics[5], biology[6, 7], nonlinear optics[8, 9], Bose-Einstein condensates[10], plasmas[11], and other physical settings. Vector SP (VSP)[12-14] represents a class of a pulsating two-component solitons in which coherently interacting components demonstrate periodic energy exchange. In passive mode-locked fiber lasers (MLFLs), SP generally exists in short-period and long-period pulsations (SPP and LPP). They appear, respectively, outside of the stationary mode-locking region[15], or when the pump power is reduced below the level maintaining the stationary mode-locking regime[16, 17]. Besides, the combination of SPP and LPP produces more complex and less stable multi-period VSPs[18]. Single-SP[19-23] and multi-SP[24-26] regimes in MLFLs were chiefly studied by numerical simulations in theory and time-stretching dispersion-Fourier-stretching technique (TS-DFT) in experiments. The transition from stationary solitons[26] to dynamical chaos is related to the sudden change of VSP[18, 27-29] period caused by the higher-order dispersion and nonlinearity. Understanding dynamics of complex VSP in highly nonlinear nonstationary soliton states is essential for the studies of dynamical chaos, spectral analysis, fiber-optic communications.

---

[#] These authors contributed equally to this work.

[*] Corresponding author email: liuwei@zafu.edu.cn; wangyy424@163.com; dcq424@126.com

Modeling femtosecond MLFLs by means of numerical simulations[16, 30] requires to accurately map parameters of the complex cavity into those of the respective nonlinear Schrödinger(NLS) system. Output states corresponding to nonstationary solutions are especially sensitive to small variations of input parameters, the efficiency of the respective split-step Fourier method (SSFM) used for the simulations being quite low. For the reverse design and optimization of the ultrafast fiber laser, it is necessary to further debug and run simulations with different parameters, which SSFM allows to perform with a notoriously low time utilization rate. Compared with the traditional numerical methods, the deep learning (DL)algorithm[31-35] is efficient for constructing complex dynamical models based on the system's parameter optimization and observation data. This technique was successfully applied to ultrafast photonics and other highly nonlinear optical settings[36-38] to predict better performance and find wider applications. The physical informed neural-network method (PINN) is used to predict soliton dynamics in optical fibers[39]. However, because of the strong correlation with the modeled physical equations, PINN is not suitable for the prediction of the long-distance evolution of optical signals in MLFL involving complex cavity parameters.

Considering the extreme complexity of the temporal and spectral evolution, recurrent neural networks (RNNs)[40-42] have attracted attention due to their vast internal memory, which allows one to model the nonstationary soliton dynamics and predict the soliton evolution. Recently, Lauri *et al.* used a long short-term memory (LSTM) RNN to predict the dynamics of the complex formation of ultra-wideband supercontinuum in optical fibers[43]. However, this technique is not yet applied to the soliton dynamics in MLFLs, especially as concerns the nonstationary dynamics. In MLFLs, algorithms such as the feedforward neural network[44] and residual convolution neural network[45] have been used to analyse the spectral interference of pulses and detuning of steady-state modes. However, due to the complexity of the system's parameter matching in MLFLs and high accuracy of the experimental data collection, the results predicted by the two above-mentioned neural networks were not, as yet, consistently compared with experimental findings, which reduces practical possibilities for applications of the networks. Besides that, RNN can effectively predict the transmission of the signal in the optical fiber, hence it offers a basic application to fiber-laser information coding[46,47]. RNN rapidly and accurately predicts the signal output based on its internal memory, which reduces the impact of the design cost and output performance instability of the laser in the information-coding system. Therefore, using the advantages of the RNN model in the context of laser optics has become an emerging topic in the current research work.

In this work, we adopt two-parallel bidirectional LSTM(TP-Bi_LSTM) RNN with the attention mechanism[48] to improve the prediction efficiency of two-component vector signals. Predictions made in this work are not based on the input of the laser's work parameters to optimize the laser cavity or to a priori predict sequences that may emerge from it, which has been well realized by the genetic[49] and human-like algorithms[50], etc. Instead, the predictions are based on a sequence in the situation when the laser is already operational. The work successfully predicts the subsequent evolution of the signal from non-stationary soliton dynamics solely by relying on a small amount of initial data with the high accuracy, greatly reducing calculation and time costs. The predicted results are consistent with the transient information obtained by means of the TS-DFT. Moreover, the successful prediction of VSP regimes for stable and unstable states indicates that the TP-Bi LSTM RNN technique developed in this work can predict the complex sequence behaviour of solitons with excellent data-analysis and forecast abilities.

Furthermore, the scheme of replacing the pulse signal by the predicted result of RNN for coding information storage is confirmed, and a coding system based on different VSPs is realized by reasonably using the predicted pulse signals verified by the experiment. This work not only provides a novel approach to the modelling of the nonstationary soliton dynamical system, but also proposes a new way for the DL prediction in the VSPs studies and offers an effective scheme for guiding ultrafast optical applications by neural network.

## 2. Results
### 2.1 Procedure

In the complete description of this work provided in Fig. 1, the RNN does not completely replace the use of numerical simulations in Part I. Actually, RNN models with supervised mechanism in Part II require datasets as the modeling basis, which is provided by numerical simulations of coupled NLS equations (CNLSE) in Part I. Different cavity parameters (see Appendix I) in the complex experimental system of Part III are used in the simulations to make the training sets highly reliable. Our method cannot completely substitute numerical simulations because the required samples of training sets should be obtained by adding a perturbation of the relative size of $10^{-6}$ - $10^{-8}$ to the group velocity dispersion coefficient in the gain fiber in CNLSE, which is difficult to implement in the experiment. Once the RNN modeling is successful, rapid and accurate prediction of subsequent pulse trains can be achieved, based on a small number of initial pulses. Then, the predicted results in part II are compared with experimental results captured by the high-resolution optical equipment in part III, to judge the accuracy of the prediction. In the experiment with parameters given in detail in Appendix III, the mode-locking is provided by saturable absorbers (SAs) based on carbon nanotubes (CNT). Further, based on the the predicted pulse signals verified by the experiments, the coding information storage system is realized in Part IV.

For the RNN employed in Part II, the TP-Bi_LSTM RNN structure, which is shown in Fig. 2, combines the forward and backward information concerning the input sequence, and introduces the attention mechanism to improve the prediction ability in terms of the overall perspective. Considering two polarization directions of VSs, a dual-signal input structure is introduced to realize the parallel input and parallel output of two vector signals. The training set of RNN implemented in the present work uses 500 samples of VSPs with 1500 round trips (RTs) provided by CNLSE, of which 450 and 50 samples are used for training and testing, respectively. This approach avoids problems of large calculation and high time cost encountered by numerical simulations and DL algorithm for MLFL. RMSprop optimizer with adaptive learning rate is used to optimize the training initial data set. In the course of the training process, through a large number of labeled data ($X_s$, $U(z+s\Delta z,t)$, $V(z+s\Delta z,t)$) with $s = 1,2,\cdots l$, RNN learns the loss function $\Gamma_\zeta$ defined by the weights and deviations $\zeta \in \xi$ of the neural network, and finds the optimal solution $\Gamma_{\zeta^*}$. Here, $X_s$ is the $s$-th pulse input, $U(z+s\Delta z,t)$ and $V(z+s\Delta z,t)$ are pulse envelopes of the two vector components at different positions, $\xi$ is set of all weights and deviations in the neural network. $\Gamma_\zeta$ and $\Gamma_{\zeta^*}$ are calculated as

$$\Gamma_\zeta = \frac{\sum_{m,h,l}(\psi_{m,h,l} - \hat{\psi}_{m,h,l})^2}{\sum_{m,h,l}(\psi_{m,h,l})^2} \qquad (1)$$

$$\Gamma_{\zeta^*} = \underset{\zeta \in \xi}{\mathrm{argmin}} \|\Gamma_\zeta\| \qquad (2)$$

where $\psi_m$ and $\hat{\psi}_m$ are the pulse envelopes with components $U$ and $V$ of the $m$-th sample, as predicted by RNN and simulated by the CNLSE respectively, $h$ and $l$ being the sets of the energy distribution of the optical pulses and numbers of the evolution RTs, respectively.

To test the ability of RNN to predict the generalization of various VSPs, we let it learn solely transient changes of 0-1000 RTs in the training set, and then use the pulse intensity of the first cycle of 50 samples in the test set to predict the dynamical set of VSPs with 500 RTs more than were used in the test set, without considering the influence of initial parameters of the system on the RNN. Detailed information about TP-Bi_LSTM RNN is given in Appendix II.

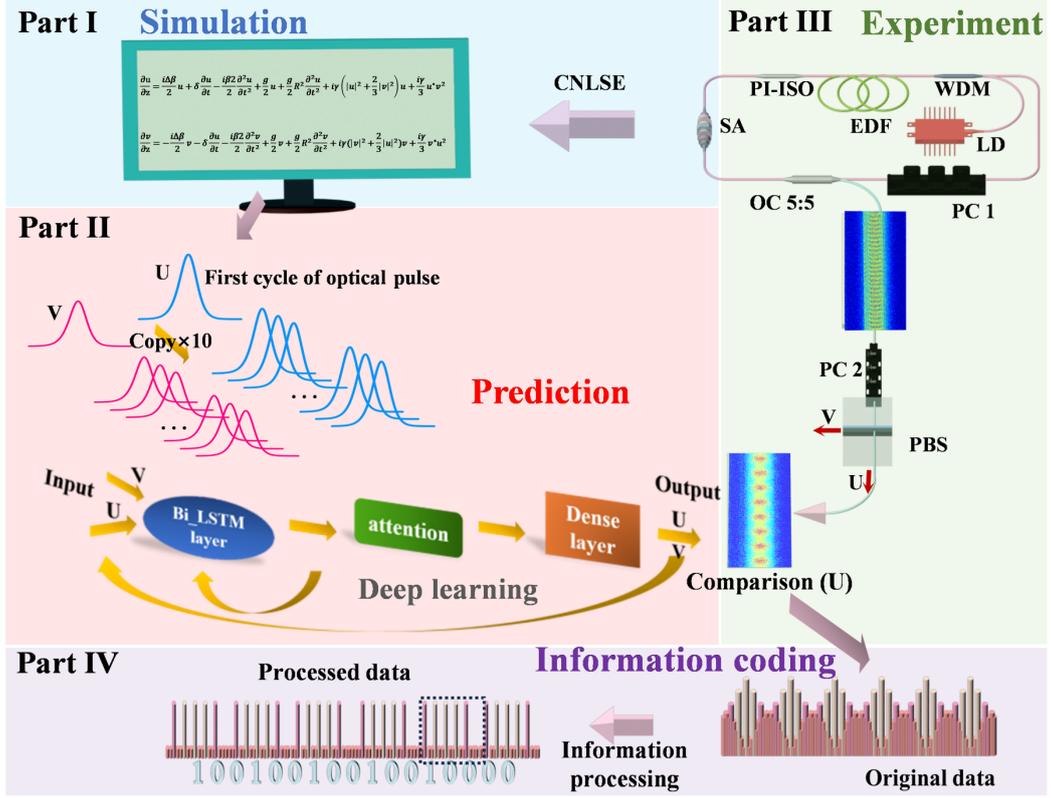

Fig. 1 The flow charts of the RNN prediction for the soliton dynamics and coding information storage, and the experimental setup.

We measure the accuracy of the RNN prediction in two forms, namely, the error and root-mean-square error (RMSE), as

$$\text{Error} = x_{m,h} - \hat{x}_{m,h}, \quad (3)$$

$$\text{RMSE} = \sqrt{\frac{\sum_{h,l}(x_{m,h,l} - \hat{x}_{m,h,l})^2}{\sum_{h,l}(x_{m,h,l})^2}} \quad (4)$$

Where $x_m$ and $\hat{x}_m$ are the pulse envelope energies of the *m*-th sample, as predicted by the RNN and measured by the experiment, respectively, and the *x*, *m*, *h*, *l* mean exactly what is defined in the loss function.

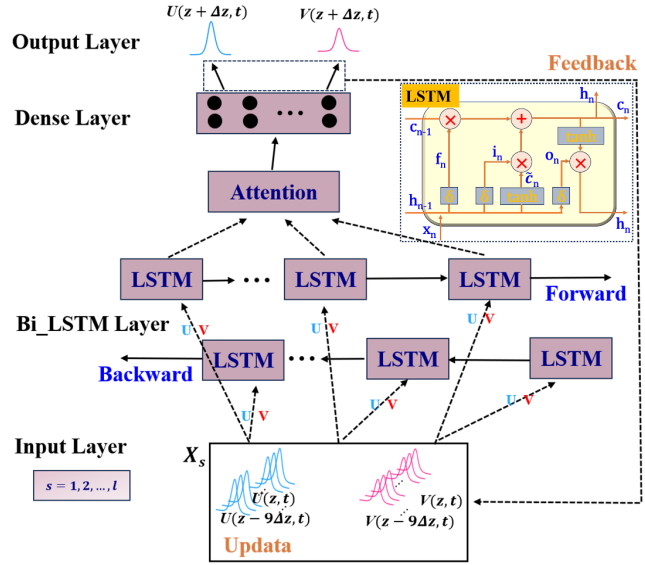

Fig.2 The TP-Bi_LSTM RNN with the attention mechanism. The realization of the data-driven pulse evolution includes the input, TP-Bi_LSTM, attention, density, and output layers.

## 2.2 The prediction analysis

*2.2.1 The dynamical prediction and experimental verification of the periodic VSP*

We train the RNN to predict a periodic VSP regime in the vector-soliton state. Periodic VSPs emerge above the threshold of the modulational instability in the MLFL with the average anomalous dispersion. The change of the system's parameters leads to a difference in the pulse's peak power, energy and pulse duration during the transient evolution of VSP. Fig. 3(a) demonstrates accurate match between the predicted and experimental results for the periodic VSPs -- for example, for the period-21 with two orthogonal components. VSP with period-21 means that the pulse returns to the initial state after passing the distance corresponding to 21 RTs. Consistent results from the prediction, simulations, and experiment for the VSP in the course of 130 RTs in Fig. 3(b) hint that single bell-shaped profiles of the solitons underlie the periodic energy evolution of two orthogonal components $U$ and $V$ with the energies

$$E_U = \int_{-\infty}^{+\infty} |U(z,t)|^2 \, dt, \ E_V = \int_{-\infty}^{+\infty} |V(z,t)|^2 \, dt \qquad (5)$$

.

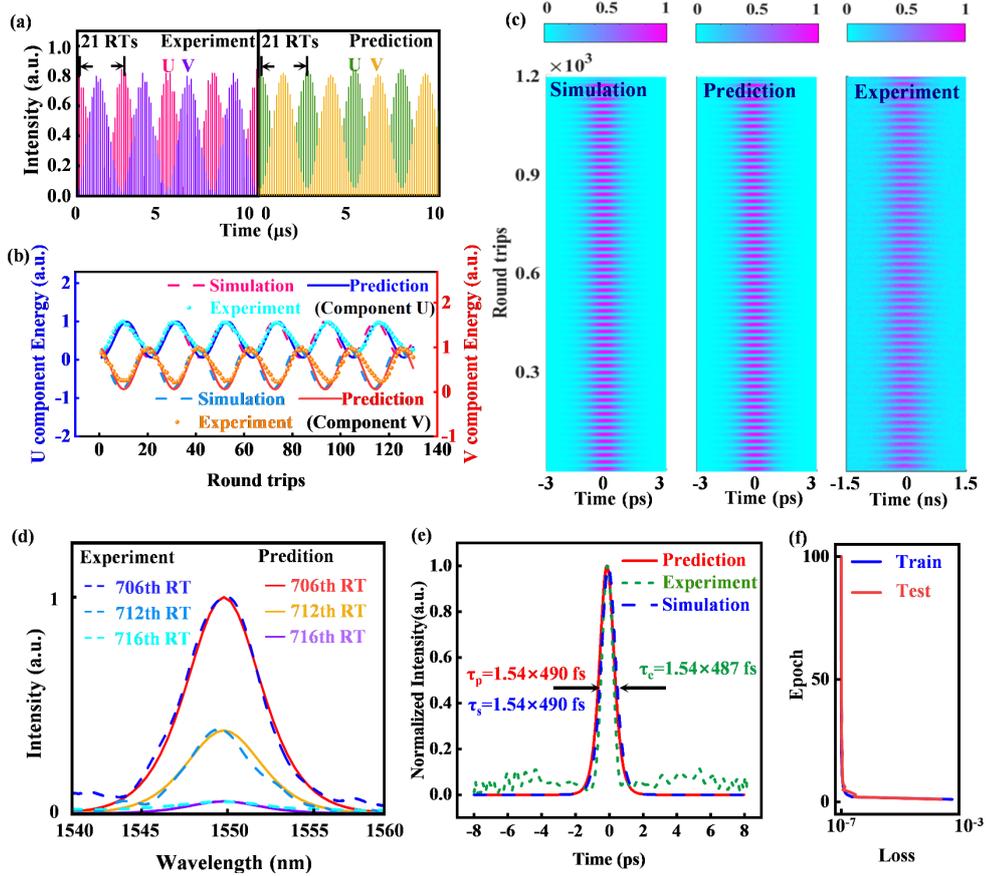

Fig. 3 VSPs with the period-21. The pulse sequences for components U and V, as produced by the prediction and experiment are plotted, in panels (a). Panel (b) displays the comparison of the prediction, simulations, and experiment for the energy envelope during 130 RTs. (c) The VSP dynamics in the course of 1200 RTs. (d) The comparison between the predicted and experimentally observed transient spectra at 706th, 712th and 716th RTs. (e) The pulse's width, as obtained from the simulations, prediction, and experiment. (f) Loss function iterates with epoch.

The energy evolution is analysed by fitting the observed VSPs to a simple harmonic form, $A(t) = A_0 \cos(\omega t + \varphi)$, with amplitude $A_0$, frequency $\omega$, and initial phase $\varphi$. The instance, the comparison of the predicted harmonic form, $A_p(t) = 0.47\cos(\frac{\pi t}{10.45} + 0.98\pi)$, and its experimentally found counterpart, $A_e(t) = 0.46\cos(\frac{\pi t}{10.65} + 0.97\pi)$ shows close proximity between them. In addition, it is relevant to stress the synchrony of the evolution of the $V$ and $U$ components. In particular, the phases constants in component $V$, corresponding to the prediction and experiment, are, $1.97\pi$ and $1.92\pi$, respectively. The initial phase shift between components $U$ and $V$ is $\pi$, making the SPs of two components asynchronous, namely, the pulse intensity peak of component $U$ corresponds to the intensity valley of component $V$ in each period, and vice versa. The RMSEs of components $U$ and $V$ between the data generated by the prediction and the experiment are 0.0299 and 0.0396, respectively. The same for the discrepancy between the numerical simulations and experiment are 0.0469 and 0.0454 for the $U$ and $V$ components, respectively. Deviations from the experiment, produced by the RNN predictions in components $U$ and $V$, are, respectively, 36% and 13% less than the deviations of the CNLSE simulations, which indicates that the RNN matches the experiment essentially better than the traditional SSFM. In the following, we focus on the dynamics of component $U$ with the larger magnitude.

The VSP dynamics with period-21 for component $U$, as produced by the prediction, simulations, and experiment in Fig.3(c) demonstrates an obvious envelope structure featuring periods containing 21 RTs, with a large modulation amplitude, as confirmed by the comparison of the energy evolution. The purpose of the use of TS-DFT in the experiment is to map the spectral information of the pulse onto the time domain by using the dispersion-broadening effect, so as to measure the transient characteristics on the temporal scale of nanoseconds. In Fig.3(c) we compare the corresponding results as obtained from the prediction, simulations, and experiment. In addition, a typical transient spectrum in VSP is predicted and compared with the experiment. As shown in Fig. 3(d), the normalized spectra for maximum, median and minimum amplitudes, corresponding to 706th, 712th and 716th RTs, respectively, demonstrate close agreement between the prediction and experiment. To illustrate the reliability of the RNN prediction, the pulse's width is compared as a measure of the time-domain structure. In Fig. 3(e), the pulse width in the experiment, measured by the autocorrelation instrument, is $\tau_e=1.54\times487$ fs, where the original curve is 1.54 times that of the hyperbolic secant ($sech^2$) fitting curve. The pulse widths $\tau_p$ and $\tau_s$ produced by the RNN prediction and CNLSE simulations are both $1.54\times490$ fs, which is quite close to the experimental results and indicates that the RNN provides remarkable accuracy in predicting the periodic VSPs. In Fig. 3(f), the loss function for training and testing gradually converges, with the accuracy $\sim 10^{-7}$, with the increase of the number of epochs, which indicates that there is no overfitting or underfitting during the RNN operation.

In addition, we have also explored examples of the periodic VSP, such as ones with period-18 and period-39 (see Appendix IV), which also demonstrate accurate results produced by the prediction.

*2.2.2 The dynamical prediction and experiment verification of the bi-periodic VSP*

Next, we address the efficiency of the RNN in predicting bi-periodic VSP combined with LPP and SPP. Fig.4 shows the comparison of the results of the RNN, simulations, and experiment in the studies of the dynamics of bi-periodic VSP, using an example of the complex VSP with the combination of period-3 and period-43. VSP with the combination of period-3 and period-43 implies a pulse sequence including 43 RTs of LPP and 3 RTs of SPP. With the increase of the RTs in Figs.4(a)-(c), the modulation instability, caused by the interaction between the nonlinearity and dispersion effect in the cavity, breaks the stability of the SPP with the period-3 to form LPP, and then produces a stable oscillation state.

The radio-frequency (RF) spectrum in Fig. 4(e) is the proof of VSP, as obtained from the experiment and prediction, and in the time-domain the same is displayed in Fig. 4(d). The experimental results, plotted by the red dotted line, indicate that the frequency intervals of 5.86 MHz and 0.4 MHz are, respectively, about 1/3 and 1/43 of the fundamental repetition frequency, which verifies the combination of period-3 and period-43 of the VSP. The predicted results for the secondary-peak's spacing of 5.88 MHz and 0.4 MHz [the green solid line in Fig. 4(e)] confirm accurate agreement with the experimental findings. The comparison of the spectral data obtained from the 1000th RT between the RNN prediction and experiment, which is displayed in Fig. 4(f), exhibits good agreement, which confirms the reliability of the prediction of the spectrum in the highly nonlinear system. The energy changes during 150 RTs, as produced by the prediction, simulations, and experiment in Fig. 4(g), indicate that obvious energy oscillations appear between the LPP and SPP for the bi-periodic VSP, which is different from the single bell-shaped profile of the periodic VSP in Fig. 3(c).

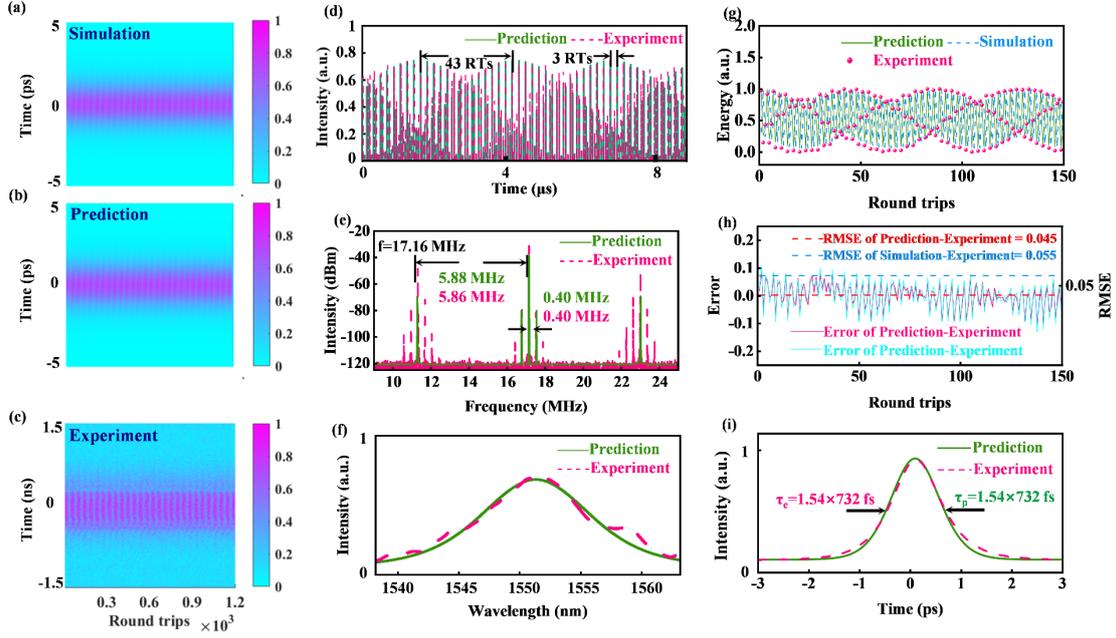

Fig. 4. The bi-periodic VSP with the combination of period-3 and period-43, as demonstrated by the simulations, prediction, and experiment. (a)-(c) The evolution picture, (d) the pulse sequence, (e) the RF spectra, (f) the comparison of the spectra between the RNN and experiment during the 1000th RT, (g) the energy envelope, (h) the pulse signal error and RMSE during 150 RTs, and (i) the pulse's width, as obtained from the prediction and experiment.

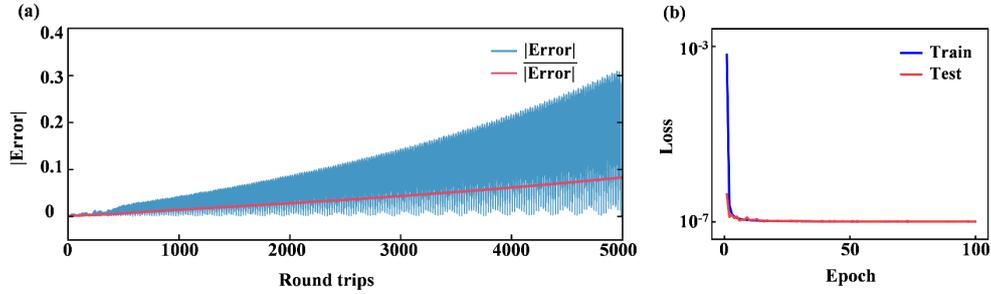

Fig.5 (a) |Error| and $\overline{|Error|}$ in 5000 RTs and (b) Loss function iterates with epoch of the combination of period-3 and period-43.

The experimental data as criteria are used to compare the RNN prediction and CNLSE simulations. Fig. 4(h) shows that both methods give rise to errors fluctuating in the range of 0.1, but the error of the oscillation amplitude, obtained from the prediction, is smaller than its counterparts obtained from the simulations. Moreover, the RMSE values between the results obtained from the two theoretical methods and experimental data are, respectively, 0.045 and 0.055, which implies high accuracy. The RNN prediction produce the deviation which is 18% less than the same produced by the CNLSE simulations, which shows that the prediction accuracy of the RNN for the VSP with the combination of SPP and LPP is still better than that provided by the simulations. Further, the pulse widths derived from the prediction and obtained from the experiment in Fig. 4(i) are both 1.54×732 fs, which proves that RNN is still credible for treating the in bi-periodic VSP.

To demonstrate the excellent predictive ability of this RNN, we use a composite VSP with the combination of period-3 and period-43 as an example, to predict the evolution of VSP in the course of 5000 RTs, which is four times larger than 1000 RTs of data used in the training set. Comparing the predicted results with the training sets, the error and mean absolute error are shown in Fig. 5(a). The error oscillates continuously throughout the entire process due to the periodic changes of the VSP intensity. The fluctuation range of the predicted error remains within 0.2 with the mean absolute error of 0.059 for the first 3900 RTs. Then the fluctuation range of the error increases after 3900 RTs, and the error attains 0.31 at the 5000th RT, the mean absolute error being 0.082 for the first 5000 RTs. These results demonstrate that TP-Bi-LSTM achieves reasonable accuracy in predicting results for the propagation distances which are several times longer than the training length, while the mean absolute error remains at a reasonable level, demonstrating the excellent prediction ability of RNN. The loss function curve of RNN for the combination of period-3 and period-43 VSP in Fig. 5(b) shows that the loss function gradually converges to accuracy ~$10^{-7}$ during the training process of the bi-periodic VSP, which indicates that the training ability of RNN for the VSP with the complex structures is still excellent.

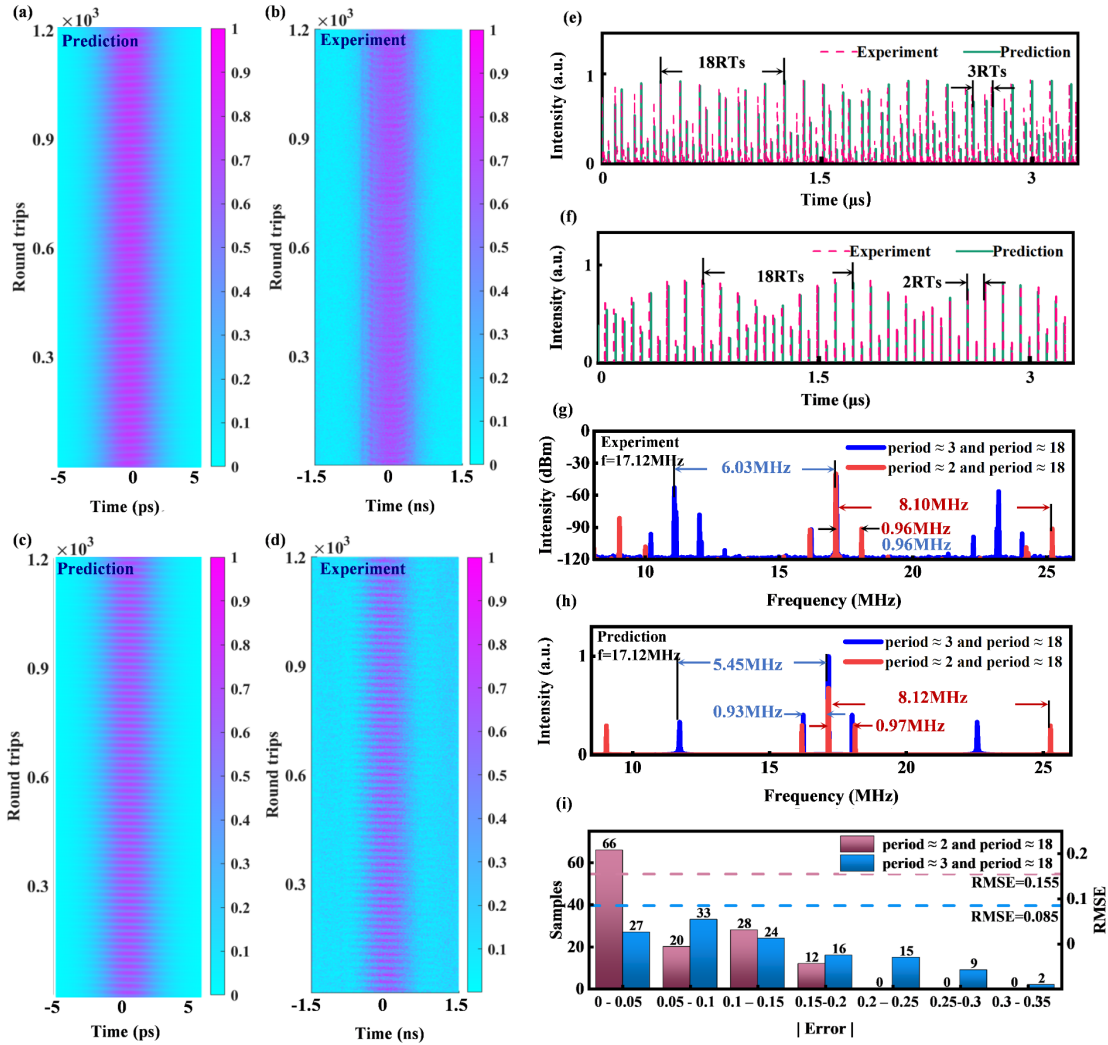

Fig. 6 Results for other bi-periodic VSPs, as obtained from the prediction and experiment. Shown

are the results for the VSPs with (a,b) the combination of period-3 and period-18, and (c,d) the combination of period-2 and period-18. The pulse sequences are shown for the combination of period-3 and period-18 in (e), and for the combination of period-2 and period-18 in (f). Panels (g,h) and (i) display severally the RF spectra and error-RMSE distributions for the two present examples of VSPs.

To prove the universality of the prediction effect of the RNN model, we study VSP with more diverse combinations of LPP and SPP, namely, bi-periodic VSPs with the combination of period-3 and period-18 and the combination of period-2 and period-18. In Figs. 6(a)-6(f), the evolution patterns and pulse sequences of these examples of VSPs show good agreement between the results of the RNN prediction and experiment. Figs. 6(g) and 6(h) display good match between the RNN prediction and experiment for the frequency interval between the side peaks and repetition frequency peak of the bi-periodic VSPs, *viz.*, ≈ '1/2 and 1/18' and '1/3 and 1/18' of the fundamental repetition frequency. In Fig. 6(i), the entire error of the VSP with the combination of period-2 and period-18 and 80% of the error of the VSP with the combination of period-3 and period-18 are basically within the range of 0.2 for all samples of 127 RTs, their RMSEs being 0.085 and 0.155, respectively. These results imply that the TP-BiLSTM RNN shows excellent data-analysis ability and prediction power for the bi-periodic VSP. Besides that, we also add an unstable case of VSP in Appendix IV to exhibit the ability of RNN to reliably deal with random events.

### 2.2.3 Unstable state of VSP

An unstable case is added to verify the prediction effect of RNN more comprehensively. Figs. 7(a,b) show the reconstruction of VSP with the period-9 within 400 RTs, as produced by the prediction and experiment, in which the intensity of the pulse gradually rises to the stable state, and the excessive SPM leads to a nonlinear phase shift and time shift of the soliton to the right. From the rising curve of the peak power between the prediction and experiment in Figs. 7(a,b) and 7(c), corresponding to the formation stage, the VSP experiences fierce competition between the SPM and dispersion, and its oscillation period varies between 11 RTs and 9 RTs.

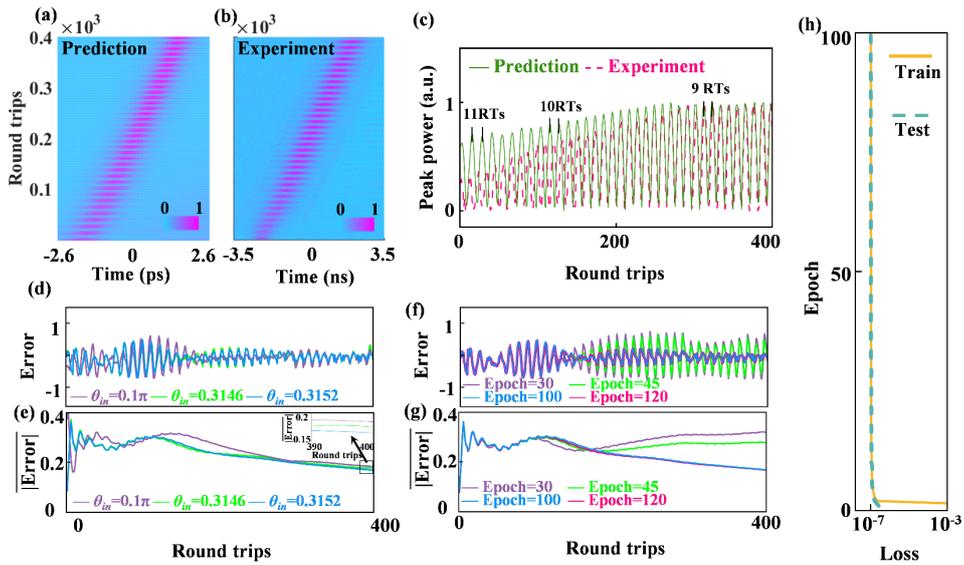

Fig.7 The unstable VSP state of with period-9. Panels (a,b) show the random reconstruction process of VSP. (c) The peak predicted and experimentally found peak power in the unstable state. (d) The error and (e) the mean absolute error for $\theta_{in}$=0.1π, 0.3146, 0.3152. (f) The error and (g) the mean absolute error for Epoch=30, 45, 100, 120. (h) Iterations of the loss function with the growth of the epoch in the unstable VSP state.

When VSP is in an unstable state, it is difficult to determine the complex relationship between the actual SPM and dispersion in the cavity. To find ways for the reduction of the prediction errors and improvement of the prediction accuracy of VSP in an unstable state, we compare the impact of training sets on prediction errors and the mean absolute error for different PC angles $\theta_{in}$ in Figs. 7(d) and 7(e). From here it is concluded that the RT error of the prediction results has a relative minimum at $\theta_{in}$ = 0.3152, which indicates that the choice of the suitable training set is crucial for reducing the prediction error of the prediction model.

Moreover, in Fig. 7(f), at $\theta_{in}$ = 0.3152, the predicted RT error gradually decreases when the number of training epochs grows from 30 to 45, to 100, and finally to 120. By analyzing the RT error in Fig. 7(f) and the mean absolute error in Fig. 7(g) for $\theta_{in}$ = 0.3152 and epoch = 100, the RT error corresponding to the initial VSP oscillation period changes between ± 0.5, while the mean absolute error is above 0.2. When SPM is balanced by the dispersion effect, a stable VSP with period-9 is obtained, and its RT error is less than 0.1, while the mean absolute error falls below 0.17. These results indicate that optimizing the training set and increasing the number of training iterations can improve the prediction performance of TP Bi-LSTM, even for unstable VSP states. In Fig. 7(h), the train-test curves of the loss function still converge, with accuracy ~$10^{-7}$, which indicates that the training ability of RNN for the unstable state of VSP is still excellent.

Table 1. The comparison of CPU and RAM resources between the CNLSE simulations and RNN prediction

| | CGLE/RNN | | | | | |
|---|---|---|---|---|---|---|
| | Period = 18 | Period = 21 | Period = 39 | Period = 3 and Period = 43 | Period = 3 and Period = 18 | Period = 2 and Period = 18 |
| Points | 4096/512 | 4096/512 | 4096/512 | 4096/512 | 4096/512 | 4096/512 |
| Steps | 1500/1500 | 1500/1500 | 1500/1500 | 1500/1500 | 1500/1500 | 1500/1500 |
| Real | 500/500 | 500/500 | 500/500 | 500/500 | 500/500 | 500/500 |
| CPU | 200%/85% | 200%/84% | 200%/81% | 200%/90% | 200%/90% | 200%/88% |
| RAM | 2.81GB/0.99GB | 2.81GB/1.17GB | 2.81GB/0.96GB | 2.83GB/1.01GB | 2.78GB/1.03GB | 2.86GB/0.98GB |
| Training time | N/A/25h | N/A/24h | N/A/25h | N/A/25h | N/A/24h | N/A/25h |
| Simulation time | 45h/95s | 45h/93s | 46h/91s | 45h/99s | 45h/98s | 46h/95s |

All the code developed based on Python 3.5 and MATLAB v2018a.
All the numerical examples reported here are run on a server with a 2.30-GHz 2-core E5-2697R processor and 256 GB of memory.

The TP-Bi_LSTM RNN with an attention mechanism is an appropriate alternative to the traditional numerical simulations for obtaining predicted results more consistent with experiment findings. The number of grid points greatly influences the accuracy and speed of the numerical simulations, while this problem can be effectively solved by the RNN. The comparison of computer resources consumed by the CNLSE simulations and RNN prediction in Table 1 indicates that the CPU in the CNLSE simulations is always running at full load, while the dependence of the RNN on CPU is obviously reduced, which is an advantages of the RNN, offering the less critical dependence on the resources and saving the memory. Moreover, the trained RNN takes about 95 seconds to produce the simulation data which are highly consistent with the simulation results obtained in the course of dozens of hours, thus the prediction speed may be dramatically improved.

## 2.3 Coding information storage of VSP based on the RNN prediction with the experimental verification

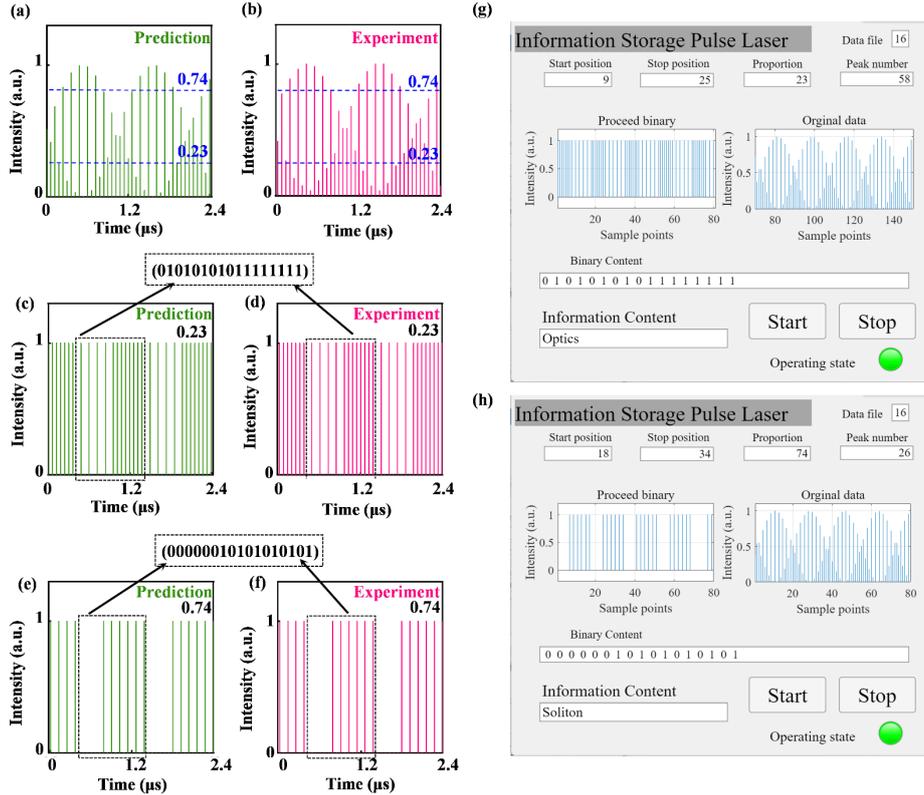

Fig.8 The information-storage setup based on the VSP with the combination of period-2 and period-18. (a) Predicted and (b) experimental results obtained during the 2.4 μs time interval. (c,d) and (e,f): Binary forms for the threshold levels of 23% and 74%, respectively. (g,h): Information-storage examples in GUI.

The signal prediction based on RNN can reduce the deficiencies, such as poor stability and slow state switching speed encountered in conventional information coding, so as to realize the fast, stable and accurate signal output. We propose a scheme of the coded-information storage by using the predicted bi-periodic VSP pulse sequence with SPP and LPP from the TP-Bi_LSTM RNN, instead of the actual pulse signal. Taking VSP with the combination of period-2 and period-18 as an example, different binary forms will appear when different judgment thresholds are selected, based on normalized predicted and experimental data, as shown in Figs. 8(a) and 8(b). Under the given threshold condition, the parts above and below the threshold are, respectively, defined as '1' and '0', thus different text information can be stored. If the signal thresholds of 23% and 74% are selected, 29 and 13 binary sequences with peak values

of '1' are obtained in the sampling range of 2.4 μs with cycle units '01010101011111111' and '00000010101010101' in Figs.8(c)-8(f), respectively.

By the information storage program, the binary coding of '01010101011111111' for 'Optics' and '00000010101010101' for 'Soliton' is successfully realized and exhibited in the Graphics User Interface (GUI) in Figs. 8(g) and 8(h).The corresponding GUI of the real-time pulse information identification system includes the switch button, pulse data start-stop position, signal judgment button, pulse peak number, normalized pulse signal on the corresponding oscilloscope, and processing the binary code according to the specific threshold. To realize and improve the ability of parsing the information, the pulse signal processing system incorporates a special information identification database, including data such as VSP period, amplitude intensity, and so on. These data can be summarized and extracted from a large number of repeated experiments and predictions, and serve for subsequent data processing.

Besides these cases, Figs. S4 and S5 also shows basic consistence between the prediction results, for coded sequences with other thresholds and other VSP pairs, and experimental results (see Appendix V). Compared with the pulse sequence obtained in the experiment, the encoded data processed by the RNN prediction is more stable and related only to the signal input of the first batch, which means that the intracavity errors of fiber lasers, such as those produced by the sensitivity of the polarization state and influence of fluctuations of the pump power, will no longer be a detrimental factor affecting the transmission signal. This conclusion demonstrates that it is feasible to use the results provided by the deep learning to replace the actual pulse signals as the data source, and offers a scheme for the application of the deep-learning predictions in ultrafast optics.

## 3. Conclusion

In conclusion, we have revealed that the RNN can bring new insights into the studies of nonstationary soliton dynamics in fiber lasers. By modeling the MLFL, TP-Bi_LSTM RNN can learn the complex propagation dynamics related to the periodic VSP and multi-period VSP combined with SPP and LPP in the optical fibers. The existence of the predicted VSPs is verified by the real-time transient dynamics in the experiment. Comparing the results of the RNN prediction and CNLSE simulations with the experiment, we conclude that the RMSE value of the RNN prediction is smaller than in the case of the CNLSE simulations. For the periodic VSP with the period-21 and the bi-periodic VSP with the combination of period-3 and period-43, the prediction results are better than provided by direct simulations – namely, deviations produced by the RNN results are 36% and 18% less than those provided by the simulations, respectively. All errors of the VSP with the combination of period-2 and period-18 and 80% of the error of the VSP with the combination of period-2 and period-18 are basically within the range of 0.2 for all samples based on 127 RTs. In addition, taking the VSP with the combination of period-2 and period-18 as an example, the VSP signal predicted by the RNN is consistent with the actual pulse signal in the experiment. Besides, the prediction results for the unstable VSP state with period-9 indicate that choosing a suitable training set and increasing the number of training iterations can improve prediction accuracy.

On the basis of the predicted RNN signal, a real-time pulse coding system with the simple storage mode is designed. The use of the predicted signal, instead of the experimental one, to realize the coding information storage avoids the shortcomings of the losses and low stability of optical pulses in the actual transmission. Thus, the combination of the deep-learning prediction and ultrafast optical setups offers a potential to greatly improve the use of ultrafast optical lasers in optical communication and information storage.

In the future work, it will be relevant to optimize the operating parameters of the laser and realize specific soliton states by using the excellent data-processing ability and forecast abilities of the RNN model. This will improve the output efficiency of the laser and will play a guiding role for the design of precisely controllable laser sources.

**Funding.**

National Natural Science Foundation of China(Grant Nos. 12261131495, 12075210 and 12275240); the Scientific Research and Developed Fund of Zhejiang A&F University(Grant No. 2021FR0009).